# Using Immersive Virtual Reality to Enhance Social Interaction

# among Older Adults: A Multi-site Study

Saleh Kalantari
Human Centered Design, Cornell University, sk3268@cornell.edu
Tong Bill Xu
Human Centered Design, Cornell University, tx66@cornell.edu
Armin Mostafavi
Human Centered Design, Cornell University, am2492@cornell.edu
Benjamin Kim
Division of Geriatrics and Palliative Medicine, Center on Aging and Behavioral Research, Weill
Cornell Medicine,
bek4007@med.cornell.edu
Angella Lee
Human Centered Design, Cornell University, al2354@cornell.edu
Walter Boot
Department of Psychology, Florida State University, boot@psy.fsu.edu
Sara Czaja
Division of Geriatrics and Palliative Medicine, Center on Aging and Behavioral Research, Weill
Cornell Medicine,
sjc7004@med.cornell.edu

## Abstract

Research examining older adults' interactions with Virtual Reality (VR) and the impact of social

VR experiences on outcomes such as social engagement has been limited, especially among

older adults. This multi-site pilot study evaluated the feasibility and acceptability of a novel

social virtual-reality (VR) program that paired older adults from different geographic locations

(New York City, Tallahassee, and Ithaca, N.Y) who engaged in virtual travel and productive

engagement activities together. The sample included 36 individuals aged 60 and older; 25% of whom had a cognitive impairment (CI). Older adults with and without CI reported high levels of engagement in the VR environment and perceived the social VR program to be enjoyable and usable. Perceived Spatial Presence was a central driver of the positive outcomes. Most also indicated a willingness to reconnect with their VR partner in the future. The data also identified important areas for improvement in the program, such as the use of more realistic and responsive avatars, controllers with larger controls, and more time for training. Overall, these findings suggest that VR social applications may foster social engagement among older adults.

**Keywords:** Social VR; Virtual Reality; Immersion; Older Adults; Cognitive Impairment; Usability

# 1. Introduction

Many older adults experience social isolation, defined as having few social relationships or infrequent social contact with others (Wu, 2020). It is well documented that social isolation can have adverse impacts on emotional, cognitive, and physical health (Cotterell et al., 2018; Malcolm et al., 2019; National Academies of Sciences and Medicine, 2020; Nicholson, 2012; Wu, 2020). In a systematic review of studies on social isolation and loneliness, Malcolm and colleagues (2019) found that older adults are one of the populations most at risk, in part due to declines in mental and physical abilities leading to social withdraw. In recent years, the impact of the COVID-19 pandemic has contributed to a further increase in rates of social isolation among older adults, with the World Health Organization reporting that in some countries up to one-third of the older population had regular feelings of loneliness (WHO, 2021). In formulating strategies to address the issue of social isolation, researchers have emphasized that solutions work best when they are tailored to needs of individuals. Some most common socialization activities (sports events, concerts, etc.) may be unsuitable or uninteresting to many older adults, particularly in the age of COVID-19 (Fakoya et al., 2020); thus, there is a strong need for new kinds of social programs tailored toward this population.

Of particular concern in the current study is the cyclical relationship between social isolation and cognitive impairment (CI) in older populations. CI is an umbrella term that covers many types of decline in cognitive acuity as we age, ranging from memory loss to problems with spatial orientation and judgement, among other impairments. Experiences of CI have been shown to be an important driver of social isolation in older adults, as CI can lead to the avoidance of interactions with others (Malcolm et al., 2019). At the same time, there is strong evidence that social isolation can negatively impact cognitive abilities, creating a vicious cycle (Cotterell et al., 2018; Cudjoe et al., 2020; Malcolm et al., 2019; Wu, 2020). Thus, there are important research opportunities to develop interventions that simultaneously address both CI and social isolation (Kalantari et al., 2022a). For example. this type of intervention approach, which is adopted in the current study, may merge established cognitive training strategies such as puzzles and memory games with positive social engagement opportunities (Simons et al., 2016; Souders et al., 2017; Zelinski & Reyes, 2009).

Another notable direction for supporting the social needs of older adults is technology-based interventions. Common platforms such as social media applications (e.g., email, videoconferencing), when carefully designed and implemented, have been shown to be an effective means of decreasing feelings of isolation among older adults and helping them to feel connected to friends and family (Czaja et al., 2018; Sen et al., 2021). The desire for social interaction has also been identified as a primary motivation among older adults who participate in online virtual experiences (Siriaraya et al., 2014; Siriaraya & Ang, 2012). In the realm of CI interventions, technology is playing an increasing role through activities such as computerized cognitive training (Hill et al., 2017; Latikka et al., 2021; Liu et al., 2019). One of the tremendous advantages of such platforms is that they can respond to the specific capabilities of individual users and be customized to account for the varied preferences and proclivities of a diverse range of older adults (Bol et al., 2019; Brandt et al., 2020; Fakoya et al., 2020). In one study, Buzzi and colleagues (2019) found that customization options in digital interventions for CI helped to increase user motivation and engagement and were associated with better long-term adherence to the intervention program.

## 1.1. Prior Research on Virtual Reality for Social Engagement among Older Adults

The term "virtual reality" (VR) encompasses a wide range of technologies that enable immersive, three-dimensional environments with a strong sense of presence or "being there" in the artificial space (Dilanchian et al., 2021; Suh & Prophet, 2018). Although VR is sometimes regarded as a youth-oriented technology, research has shown that it is readily embraced by many older individuals, and that it has the potential to improve older adults' sense of engagement and wellbeing (D'Cunha et al., 2019; Dermody et al., 2020; Huygelier et al., 2019; Lee et al., 2019; Thach et al., 2020). A handful of prior studies have been conducted to evaluate the use of VR by older adults with CI, and these studies have generally found positive responses, feelings of engagement, and comfort with the technology when it was carefully introduced to participants (Appel et al., 2020; Arlati et al., 2021; Dilanchian et al., 2021; Kalantari et al., 2022a; Park et al., 2020). It is important to note that responses in these studies varied quite a bit among different individual participants, and thus VR may not be suitable for all older adults. On average, however, there is a strong body of work supporting the use of VR as an engagement and cognition-enhancing mechanism for this population.

Beyond the use of VR as a cognitive-engagement environment, the topic of *social interactions* in VR poses an additional set of issues. The emergence of social VR platforms can be traced back to long-standing aspirations toward collaborative spaces that erase geographic distance (Benford et al., 2001). Studies have found that interactions in VR spaces are perceived as being more "realistic" or more similar to in-person interactions when compared against other types of digital communications (Li et al., 2019). However, the ways in which people can use this relatively novel technology to socialize are not yet fully understood. The increasing popularity of social VR has led to an emerging research agenda focused on evaluating aspects of the mediated interactions that take place on such platforms. This includes studying design strategies for virtual social spaces (Jonas et al., 2019; McVeigh-Schultz et al., 2019; Sra et al., 2018), communication modes and interactive activities in VR (Baker et al., 2019; Freeman & Maloney, 2021; Maloney & Freeman, 2020; McVeigh-Schultz et al., 2018; Moustafa & Steed, 2018), engagement strategies for long-distance couples and families (Maloney, Freeman, & Robb, 2020; Zamanifard & Freeman, 2019), and the psychology of VR self-presentation and avatars (Blackwell et al., 2019; Freeman et al., 2020; Freeman & Maloney, 2021; Kolesnichenko et al., 2019; Nowak & Fox, 2018).

For older adults, design factors and intuitive interfaces may be particularly important in promoting the adoption of social VR technology and comfort with using it (Roberts et al., 2019; Siriaraya & Ang, 2019; Syed-Abdul et al., 2019). A topic that has repeatedly emerged in the literature is the design of avatars and full-body tracking to reflect real-world experiences (Kalantari & Neo, 2020; Darfler, Cruz-Garza, & Kalantari, 2022). Older adults generally prefer "realistic engagement" scenarios in VR that reflect ordinary environments, rather than "gamified" scenarios that bestow users with superhuman abilities to perform challenging tasks (Freeman et al., 2020; Freeman & Maloney, 2021). In one notable study, Baker and colleagues (2019) conducted workshops with older adults on social VR use and found that the key drivers of the technology's acceptance in this population were *behavioral anthropomorphism* ("the embodied avatars' ability to speak, move, and act in a human-like manner") and *translational factors* ("how VR technology translates the movements of the ageing body into the virtual environment"). Thus, the careful design these environments is crucial to meeting the specific engagement needs of older populations, which may diverge from broader gamified development trajectories in the industry.

      Williamson and colleagues (2021) found that specific spatial design factors in virtual environments can have a strong effect in enabling positive social experiences, by mediating activities such as group formation and the sense of personal space during interactions. When socialization is the primary goal of a VR environment it is extremely important to incorporate functions for body language—the ability of avatars to fluidly convey the user's gestures, facial expressions, gaze direction, and other fundamental aspects of non-verbal communication (Maloney et al., 2020; Pan & Steed, 2019; Tanenbaum et al., 2020). Data from such prior studies has provided a valuable preliminary understanding of important features of VR environments that may affect social outcomes; however, the majority of these design studies have relied on participant samples that skew toward younger adults. For the potential of VR to be realized among older adults it is imperative to collect additional information about how factors such as embodiment, presence, and engagement are experienced by older users of social VR. To date, most studies on older adults and VR have focused on non-social environments, and most studies on social VR have focused on younger adults. The current study was designed to help fill this gap by evaluating the responses of older adults to a social VR application and analyzing features of the experience that may influence adoption of and engagement with social VR applications in this population.

**1.2. Study Goals and Research Questions**

The study was conducted to evaluate the responses of older adults, with and without CI, to a novel social-engagement VR environment designed by the research team. We examined the associations between feelings of VR presence and the factors of mood, perceived technological usability, social engagement, and desire to reconnect with a conversation partner at a later date. The current study did not evaluate specific VR environmental design variables; rather, we drew from prior research to try to optimize spatial presence and embodiment in the VR, and then evaluated the correlations among user-response variables to identify the impact of perceived presence and other aspects of the VR experience in relation to social engagement outcomes. The study paired older adult participants from different geographic locations (Tallahassee, Florida; Ithaca, New York; and New York City) who had no prior social connections with each other. The participants engaged in various VR activities intended to promote conversation and collaboration, including a "travel" module, which involved creating a trip itinerary and viewing immersive videos of the selected destinations, and various memory related tasks, all conducted in the presence of a moderator. Six research questions guided the study design:

The primary research question that motivated this study was: Can VR-mediated interactions produce meaningful social experiences among older adults (*RQ1*)? We combined qualitative and quantitative approaches to addressing this question. Based on positive results of previous VR interventions with a focus on restoration and relaxation for older adults in regard to moods and attitude (e.g., Appel et al., 2020; Huygelier et al., 2019; Kalantari et al., 2022a), we decided to evaluate if a social VR experiment could also enhance mood states and attitudes in this population (*RQ2*). While the previous studies on the use of VR for older adults showed the acceptance of immersive VR in this population (e.g., Huygelier et al., 2019), we also wanted to evaluate the possibility of any negative impacts in this population, such as motion sickness or excessive cognitive workload, and to measure the participants' usability ratings for the technology (*RQ3*). Next, we wanted to evaluate if individuals who engaged in pairwise social activities would demonstrate similar responses to the VR environment as their partners (*RQ4*). We were also interested in determining if participants' affective responses to the VR activities (pleasure, arousal, and dominance measures) would be correlated with their mood states and with social outcomes (*RQ5*). Finally, based on previous studies suggesting links between perceptions of spatial presence and engagement (Barreda-Ángeles & Hartmann, 2022; McCreery et al., 2015), we sought to evaluate if spatial presence in the VR environment predict social outcomes and engagement and do other variables in this study mediate this relationship (*RQ6*).

## 2. Methods

### 2.1. Design of the Virtual Environment for Social Interaction

In designing and developing the content of our VR social program, we accounted for the prior research literature (as discussed in section 1.1) and our previous experience in developing VR testing platform (Cruz-Garza et al., 2022; Kalantari, et al., 2022b; Kalantari et al., 2022c) related to movement affordances, proxemic spacing, avatar customization, gesture and posture control for the avatars, and social activity design. The engagement tasks in our VR modules were based on McGrath's circumplex model, which is split into four quadrants: generating a variety of ideas or plans, choosing a solution, negotiating with contradicting views, and executing the revised solution (McGrath, 1984). We also used a participatory approach during the design process by directly engaging with older adults to give us feedback about the planned VR components. This involved inviting older adults to our lab (using a convenience recruiting approach) and asking them to test various interaction scenarios, which were then fine-tuned based on the feedback. The final social VR product for this study included four modules, which we labeled as Training, Introduction, Travel, and Productive Engagement.

*Training Module.* Participants' first experience in the VR environment was a training module so that they could become familiar with the technology and its navigational and interaction controls. The virtual training space consisted of a relatively small room with a pleasant floor-to-ceiling window view (Figure 1a). The room contained distinct interaction areas, highlighted in different colors, where participants could read instructional text and practice completing tasks. The skills required to navigate in different sections of VR program were practiced in the learning trial, such as creating an avatar, moving from place to place at different speeds, interacting with textual elements, and grabbing and moving objects. A moderator (one of the researchers) was also available in the training area to answer questions and provide feedback as needed.

**Figure 1.** Screenshots from the Participants' View in the VR Modules: (a) Training Module; (b) Introduction Module; (c) Travel Module; and (d–e) Productive Engagement Module

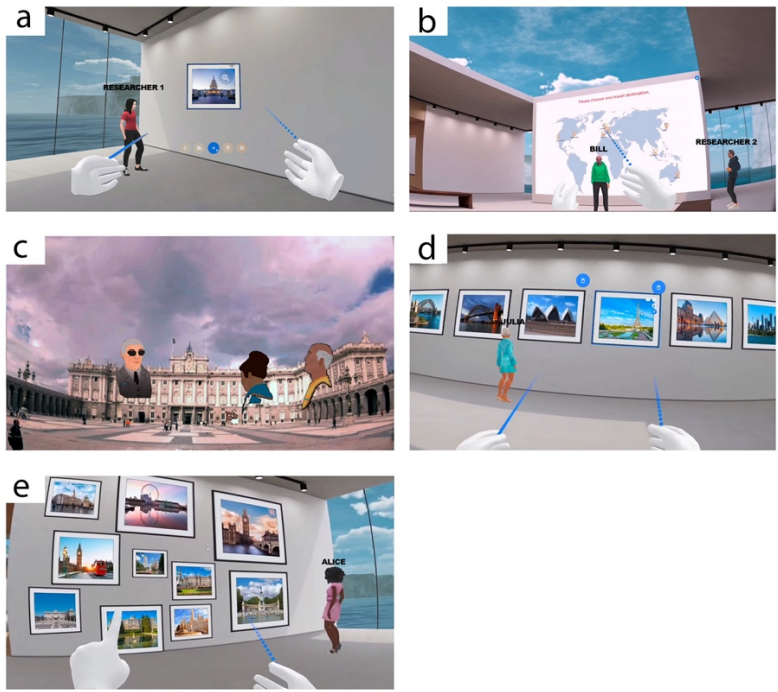

*Introduction Module.* The next area that users encountered was a foyer where they could meet other participants and learn about the social activities that were available. In the current version of the VR environment there was only one possible activity sequence, but in future work we will add additional branching possibilities. The foyer consisted of an expansive, well-lit room with a large world map in the middle (Figure 1b). Upon entering this virtual space, participants were greeted by a moderator (one of the researchers), who served to introduce two participants to each other and help "break the ice" by initiating conversations. The moderator prompted the two participants to share information about themselves such as where they were from and what they were currently doing in their lives (e.g., if they were still working or retired). The moderator then indicated the large world map and prompted discussion about past experiences with travelling and/or places that the participants might have always wanted to visit. The participants were then encouraged to agree on a location to "visit" together in the virtual experience, selecting from a wide variety of available regions or cities indicated on the map.

*Travel Module*. After agreeing on a destination, the participants engaged in the selected virtual travel experience, where they watched 360-degree videos of the location that they chose in the previous module. The participants were encouraged to discuss their experience. The moderator/researcher accompanied the participants on this journey and helped to prompt conversations about features of the environment if needed. The 360-degree videos primarily presented major landmarks and recognizable tourist destinations. Some landmarks in the videos were indicated by name through text appearing in the environment. Due to the nature of the video technology, participants were not able to voluntarily move around in these spaces or control the pace of the videos, but they could turn and look in various directions, and see each other's virtual avatars in the environment (Figure 1c).

*Productive Engagement Module*. After completing the travel segment, participants next encountered a gallery space, where they were asked to work together to complete a memory puzzle task and a creativity task. The moderator/researcher remained available for assistance if requested but did not actively participate. The gallery contained 32 photos from various recognizable world locations, including 8 photos selected from the specific videos that the participants had experienced in the previous travel module. First, the participants worked together to identify photos from their selected travel destination, and to discard irrelevant photos. They were not graded for accuracy in this task. After agreeing on a set of photos representing their virtual travels, the participants were asked to organize the photos into a collage on the walls of the room, using any kind of desired artistic arrangement (Figure 1d–e). Like the other modules described above, these activities were intended to present several dimensions of the McGrath circumplex model; in this case the selection of photographs involved memory and intellectual problem-solving (Type 3) and often resolving diverging viewpoints (Type 5), while the collage arrangement served as a creative psychomotor performance task (Type 8).

The virtual environments for the Training, Introduction, and Productive Engagement modules were developed using the Spatial platform v.6.19 (www.spatial.io). For the Travel module we used pre-made 360-degree videos available in the Alcove VR app v.1.194 (www.alcove.com). Audio connectivity was managed through the Zoom app. We chose this environmental design as the avatars have high realism. The Spatial platform can generate 3D avatar images based on an uploaded photo of the participant. For the 360-degree video segments, the researchers built custom avatars for each participant using the Oculus Quest's native avatar system. This is not as precise as Spatial, but it allowed us to roughly customize the avatar's hair, face, eyes, skin-tone, and outfit in accordance with each participant's real-life appearance. Previous literature indicates that users of social VR environments tend to care about having representative avatars, particularly hairstyle and color, as it increases their sense of immersion (Ducheneaut et al. 2009). We determined that it would be too time-consuming and potentially distracting to let participants create their own avatars, so instead we completed this task to provide a reasonable representation of each participant's real-world appearance.

## 2.2. Participants

The evaluation of the social VR program was designed as a multi-site study, encompassing participants in Ithaca, NY, New York City, NY, and Tallahassee, FL. We recruited 12 participants at each site (36 total) using a convenience sampling method. Participants aged 60 or older were invited to participate via postings on local mailing lists and phone calls to participants from other studies. The researchers contacted interested individuals via phone or e-mail to discuss compensation, exclusion criteria (epilepsy, motion sickness, medical implants), time commitment, and scheduling. The participants were then sent an informed-consent document and a demographic questionnaire, which they were asked to complete prior to the experiment session.

Once everyone's availability was obtained, the researchers paired each participant with a partner from a different city and scheduled a joint session, in which each of the partners would visit the lab site in their respective city to use the VR equipment. The final dataset includes 18 pairings in total—6 between New York City and Ithaca, 6 between New York City and Tallahassee, and 6 between Ithaca and Tallahassee—with a total of 36 participants. All participants were over 60 years of age, with an average age of 71 (SD = 5.2). The overall sample was primarily Female (72%) and White (81%). A broad range of educational backgrounds were represented. Over half of the participants had a professional degree (n=19, 53%), 5 (14%) some college education, 10 (28%) a college degree, and 2 (6%) had a Doctorate degree.

### 2.3. Study Procedures

All study procedures and questionnaires were approved by the Institutional Review Boards at all three institutions, including Cornell University, Weill Cornell Medicine (WCM), and Florida State University (FSU) prior to the research activities. During the experiment sessions there were two researchers present at each site: one was responsible for administering the questionnaires and monitoring safety issues, and the other was responsible for coordinating the technological setup and served as a moderator within the virtual environment. All three study sites used a consumer version of the Oculus Quest 2 head-mounted display and Oculus hand-held controllers. The headsets were customized with the Oculus Elite Strap (an adjustable ergonomic support). The Quest 2 headset weighs 831 grams, has 6 degrees of freedom, and uses an LCD display with a resolution of 1832x1920 per eye and 90 Hz refresh rate. All sites also used a Blue Yeti USB microphone and a Sony SRS-RA3000 speaker for transmitting sound through Zoom. We recorded the VR display video of participants using the Side Quest app (www.sidequestvr.com).

Participants completed a demographic survey remotely before arriving at the site, and then upon arrival the researchers administered the Montreal Cognitive Assessment (MoCA) MoCA and questionnaires assessing mood states and attitudes toward VR (these instruments are described in more detail in section 2.4). During this time, a photo of the participant was also taken and used to create virtual avatars in Spatial and Alcove. The researchers then helped each participant to don the VR equipment and prepare to enter the modules. Participants sat in a swivel chair or stood as desired throughout the experiment activities, and they could move around in a space that was approximately 2 m x 2 m at each site. A five-minute break was given after each module, with participants' voice communication temporarily muted between the study sites. The start of each module was synchronized in real-time by the researchers between the two sites, so that one participant would not have to wait a significant time in the VR environment for the other participant to arrive.

Completing the Training and Introduction modules took between five to ten minutes total, depending on the time required for learning the controls and for choosing a travel destination. During the break after the Introduction, Travel, and Productive Engagement modules, participants completed the Self-Assessment Scale (SAM) to measure affective responses to the VR activities. Once the Travel module began, the moderator/researcher began to gradually withdraw from involvement in the VR, encouraging the participants to engage in organic dialogue about whatever topics they wished, and to support each other with questions or issues with the controllers. For the rest of the VR session the researcher would only provide prompts if participants directly asked for their assistance or if both participants remained continuously silent for more than two minutes. Completing the Travel module took between 12 and 15 minutes (the researchers ended the videos during an organic lull in the conversation) and completing the Productive Engagement module took between 15 and 20 minutes (ending when the participants expressed satisfaction with their photo collage design). After the VR activities were finished, each participant completed several additional questionnaires, followed by a semi-structured interview about their experiences. Each participant who completed the experiment received a $50 gift card as compensation for their time.

## 2.4. Measures

Proficiency with information technology was assessed in the initial demographic instrument using two scales: the *Computer Self-efficacy Scale* (Barbeite & Weiss, 2004) and the *Mobile Device Proficiency Questionnaire* (Roque & Boot, 2018). The first scale focuses on confidence levels in relation to information technology, while the second evaluates proficiency with using various mobile devices (tablet computers and smartphones). Both use 5-point Likert scales, with higher scores indicating greater proficiency. In the same demographic survey, we asked participants to complete the *Positive and Negative Affect Schedule (PANAS)* (Watson, Clark, & Tellegen, 1988). This scale includes 10 items measuring positive affect (e.g., excited, inspired) and 10 items measuring negative affect (e.g., upset, afraid) on a scale between 10 to 60. Finally, the initial demographic instrument also included the *20-Item Short Form Health Survey (SF-20)* (Cooke et al., 1996), which briefly evaluates aspects of physical and mental health functioning.

Immediately prior to the VR sessions, participants' cognitive capabilities were evaluated using the *Montreal Cognitive Assessment* (Nasreddine et al., 2005). This instrument does not in itself confirm a diagnosis of CI, but it is commonly used as a screening tool to assess cognitive status. The assessment involves several brief written and verbal tasks measuring executive functions, memory, language, and reasoning. In the current study the assessment was used to divide the participants into those with a likely cognitive impairment and those without, using a threshold score of lower than 26 on the instrument as likely-impaired. We chose 26 as < 26 is the most commonly used cutoff for the presence of a cognitive impairment (Carson et al., 2018; Milani et al., 2018; Wong et al., 2015). None of the participants in the current study scored below 17 on this instrument to be regarded as having a moderate or severe cognitive impairment.

The mood states of each participant were assessed immediately before and immediately after the VR session, using the *Multidimensional Mood State Questionnaire* (Steyer et al., 1997). This instrument includes 30 items, each on a 6-point Likert scale, with higher scores indicating more positive mood states. The results are divided into three dimensions, including "good/bad" mood (GB), "calm/nervous" mood (CN), and "awake/tired" mood (AT). The total scores of a participant for each of these three dimensions were calculated separately, for both pre- and post-exposure to the VR environment. Immediately before and immediately after the VR session we also assessed attitudes toward VR technology, using a scale developed by Huygelier and colleagues (2019) for the specific purpose of evaluating the *Acceptance of Head-mounted Virtual Reality in Older Adults*. This instrument includes 18 items, each on a 5-point Likert scale, with higher scores indicating a more positive attitude toward the technology. The item scores items were summed to obtain a total VR-attitude measurement pre- and post-exposure. After completing each module, participants also completed a *Self-Assessment Manikin (SAM)*. This is a pictographic scale assessed affective reactions in three dimensions of pleasure (happy/unhappy), arousal (excited/bored) and dominance (in control / not in control) (Bradley & Lang, 1994). SAM was repeated after each module, and the final SAM score on each dimension was averaged.

Participants also completed The *MEC Spatial Questionnaire* (Vorderer et al., 2004) after completing the VR tasks. This instrument was used to measure experiences of presence ("being there") in the virtual environment. The MEC contains two subscales with 6 items each, including the Possible Actions subscale and the Self Location subscale. In addition, participants were asked to complete the *NASA Task Load Index* (Hart & Staveland, 1988) to evaluate perceptions of effort and frustration. This instrument includes six subscales—mental demand, physical demand, temporal demand, effort, performance, and frustration level—each on a 5-point Likert scale, with higher scores indicating greater perceived task loads. Total scores were calculated by summing the responses across all items. They were also asked to complete the *Simulator Sickness Questionnaire* (Kennedy et al., 1993) to assess experiences of "cybersickness." This instrument includes 16 items on a 4-point Likert scale, with higher scores indicating greater discomfort. The item responses were summed to get a total cybersickness score for each participant.

The level of participants' *Engagement in the VR Experience* was also assessed following the completion of the VR tasks using a 3-item, 5-point Likert scale developed by the current researchers, with higher scores indicating greater perceived engagement. The Cronbach's Alpha of 0.81 for this new instrument indicated that there was good internal consistency in the results (Appendix A). To measure the sense that other participants in the environment were "real," the participants completed the *Social Presence Scale* (Nowak & Biocca, 2003). This instrument involves responding to 6 questions using "sliders" on a scale of 0–100, with higher scores indicating greater perceived social presence as well as the *Willingness and Likeliness to Reconnect Scale* (Boothby et al., 2018) to assess positive social outcomes. This instrument includes 8 questions in a 5-point Likert scale format to evaluate interest in continuing a social relationship in the future. Higher scores were understood to indicate a more positive view of the social interaction. Finally, they completed the 4-item *Usability Metric* (Finstad, 2010). This instrument asked participants to rate the usability of the VR system on a scale of 0–100, with higher scores indicating a more intuitive and comfortable interface. A summary overview of the quantitative measurement instruments and administration sequence is presented in Figure 2.

**Figure 2.** Summary of Measurement Tools Used in the Study and Study Flow

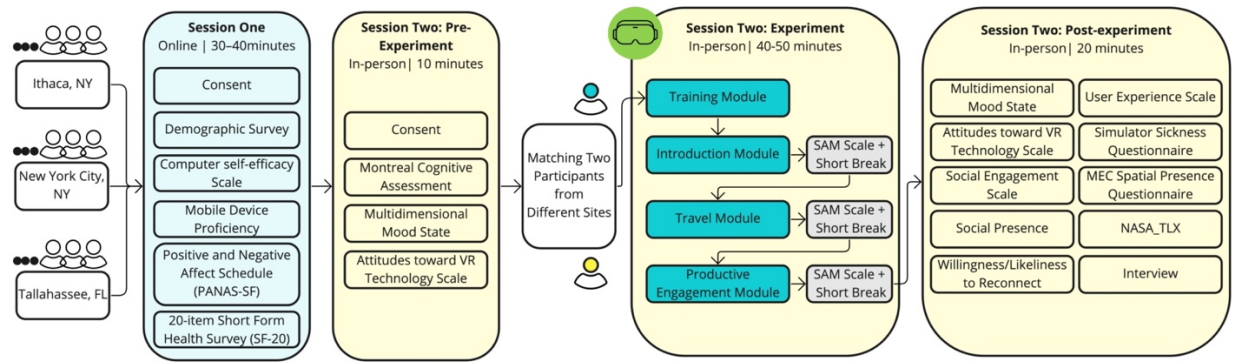

## 3. Results

We used the R programming language for statistical data analysis. In addition to conventional descriptive results, we also calculated the intra-cluster correlation coefficient ($\rho$) the library "fishmethods", following Lohr's equation no. 5.8 (Lohr, 2021, p. 139). In order to better handle the variance within each experimental pair, we fitted linear mixed models with library "lme4", then estimated and compared marginal means with Satterthwaite approximation with library "emmeans". For analyzing the enhancement of mood and attitude toward VR, we fitted linear mixed models with no fixed effect and random effects of participant pairs to estimate the change in dimensions of mood before and after the experiment and compared them with zero. We also fitted linear mixed models with fixed effects of likelihood of CI (based on a MoCA screening score with a threshold of 26) and random effects of participant pairs to estimate the differences between CI and non-CI groups in outcome variables. For testing the correlation between participants' affective responses to the VR activities with their mood states and with social outcomes, we fitted linear mixed models with fixed effects of the happy-unhappy, exited-bored, and dominance dimensions of the SAM and random effects of participant pairs to estimate the social outcomes and engagement.

While our central focus was to evaluate the impact of Spatial Presence ("being there"); we were also interested in considering mediating variables in these relationships, for which a structural equation model is a suitable approach (Hooper et al., 2008). We fitted a model to explore the underlying mechanisms behind the effects of Spatial Presence on Reconnect with library "lavaan". Note that the results to be interpreted with caution due to small sample size (n=36).

### 3.1. Participant Characteristics

As noted, the MoCA was used assess cognitive status. Our participants had an average MoCA score of 27, and 25% of the participants had a MoCA score of less than 26, indicating a likelihood of mild cognitive impairment. Participant characteristics are reported in Table 1.

**Table 1.** Participant characteristics.

| Variable | Ithaca (N=12) | New York City (N=12) | Tallahassee (N=12) | Overall (N=36) |
|---|---|---|---|---|
| **Age** | 67 (4.0) | 71 (4.8) | 74 (4.3) | 71 (5.2) |
| **Gender** | | | | |
| **Female** | 7 (58%) | 11 (92%) | 8 (67%) | 26 (72%) |
| **Male** | 5 (42%) | 0 (0%) | 4 (33%) | 9 (25%) |
| **Non-binary / third gender** | 0 (0%) | 1 (8%) | 0 (0%) | 1 (3%) |
| **Ethnicity** | | | | |
| **Other** | 3 (25%) | 1 (8%) | 0 (0%) | 4 (11%) |
| **White** | 9 (75%) | 8 (67%) | 12 (100%) | 29 (81%) |
| **Black or African American** | 0 (0%) | 3 (25%) | 0 (0%) | 3 (8%) |
| **Education** | | | | |
| **4-year degree** | 5 (42%) | 3 (25%) | 2 (17%) | 10 (28%) |
| **Doctorate** | 1 (8%) | 0 (0%) | 1 (8%) | 2 (6%) |

| Variable | Ithaca (N=12) | New York City (N=12) | Tallahassee (N=12) | Overall (N=36) |
|---|---|---|---|---|
| **Professional degree** | 5 (42%) | 8 (67%) | 6 (50%) | 19 (53%) |
| **Some college** | 1 (8%) | 1 (8%) | 3 (25%) | 5 (14%) |
| **MoCA Score** | 27 (1.8) | 26 (3.7) | 27 (1.5) | 27 (2.6) |
| **Computer Proficiency** | 28 (2.1) | 26 (3.6) | 28 (3.0) | 27 (2.9) |
| **Mobile Proficiency** | 33 (6.4) | 33 (5.1) | 34 (8.5) | 33 (6.6) |
| **PANAS** | | | | |
| **Positive Affect** | 38 (6.3) | 39 (7.1) | 34 (6.0) | 37 (6.6) |
| **Negative Affect** | 12 (3.3) | 15 (5.6) | 13 (3.8) | 13 (4.4) |
| **SF-20** | | | | |
| **Physical Functioning** | 84 (25) | 75 (30) | 75 (27) | 78 (27) |
| **Role Functioning** | 92 (29) | 79 (33) | 77 (39) | 83 (34) |
| **Social Functioning** | 88 (22) | 82 (31) | 83 (32) | 84 (28) |
| **Mental Health** | 76 (21) | 75 (13) | 77 (14) | 76 (16) |
| **Health Perceptions** | 71 (29) | 65 (29) | 70 (31) | 69 (29) |
| **Pain** | 33 (26) | 35 (28) | 38 (29) | 36 (27) |

Note: Mean (SD) for continuous variables, n (%) for categorical variables.

## 3.2. Participant Responses to the VR Environment (*RQ1–3*)

Participants reported a very high level of Engagement in the VR environment (M=4.18 out of 5.00; SD=0.91). Ratings of Social Presence were more moderate but still reasonably high (M=61.21 out of 100; SD=22.00); and the participant ratings for Likeness to Reconnect with the VR partner were also moderately high (M=3.69 out of 5; SD=0.79). The participants reported that the VR system had a high usability (M=67.01 out of 100; SD=20.73). There was some simulator sickness reported, but the rates were low (M=21.82 out of 235.62; SD=26.69). Task workload scores were also within the low range (M=2.86 out of 7; SD=1.17). Participants reported a moderately high sense of Spatial Presence in the VR, on both subscales of Possible Actions (M=3.70 out of 5; SD=1.17) and Self Location (M=3.76 out of 5; SD=1.27).

Participants felt happy (M=2.3 out of 9 reverse-coded; SD=1.4), somehow calm (M=4.7 out of 9; SD=2.1), and relatively in control (M=5.1 out of 9; SD=1.5) during the experimental session.

**Figure 3.** Distribution of Outcome Measures.

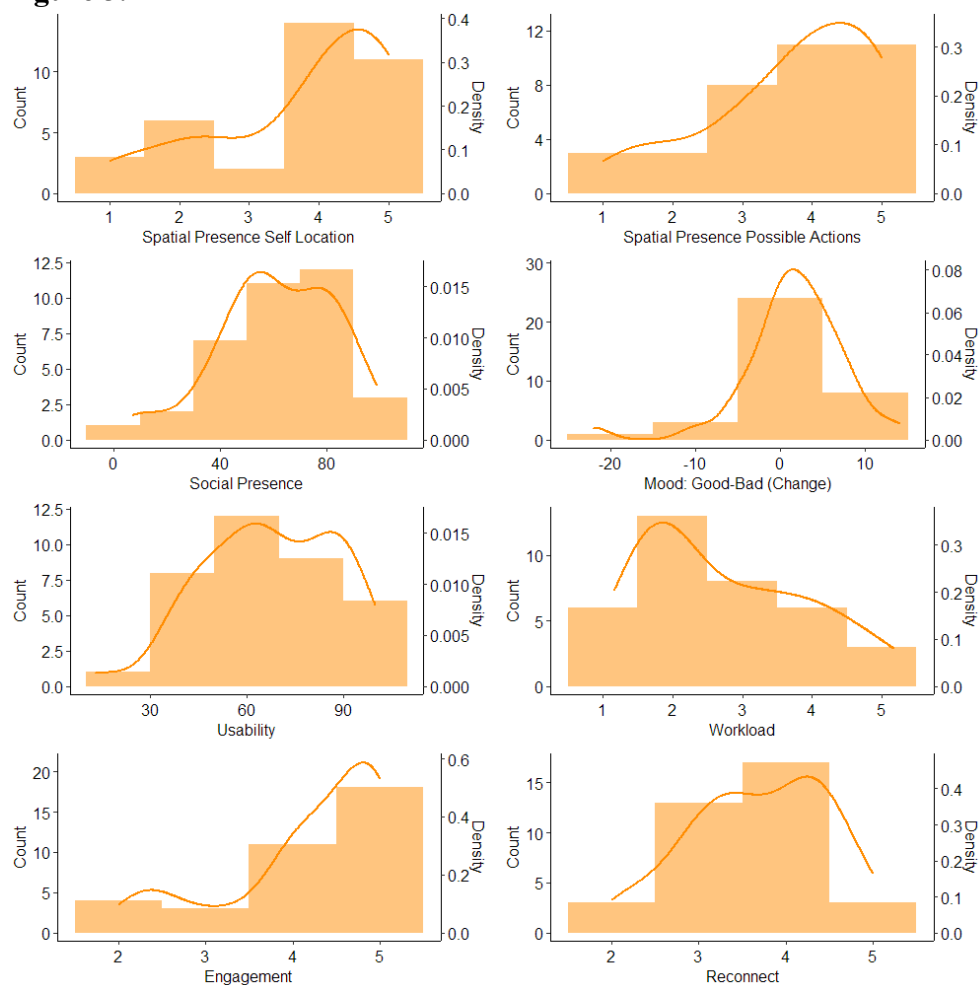

There was only a small change in participants' attitudes toward VR technology from before the experimental session to after the sessions (M=-0.05 out of [-4 to +4]; SD=0.37). Changes in mood states were small but positive, with an average 1.47-point shift toward the "good" mood state (SD=6.30), an average 3.00 point shift toward the "calm" mood state (SD=6.93), and an average 0.61 point shift toward the "awake" mood state (SD=6.61), all measured out of a theoretical range of [-60 to +60].

Only the change in the calm–nervous dimension of mood was found to be significant (t(17.0)=2.53; p=0.022; 95% CI: [0.61, 5.39]). This finding makes intuitive sense, as participants would be expected to feel more nervous at the beginning of an experiment session (Figure 4). A summary of the descriptive statistics for all outcome variables are presented in the Table 2. More information about the descriptive results for outcome variables evaluating different aspect of the social interactions including Engagement, Social Presence, and Reconnect, and descriptive results for participant affective response (SAM scale) to each module are reported in Appendix A and Appendix B.

**Figure 4.** Change in Mood from Before to After the VR Experience: Good–Bad Dimension (at Left), Awake–Tired Dimension (Middle), and Calm–Nervous Dimension (at Right); Statistically Significant Differences Were Found Only for Calm–Nervous

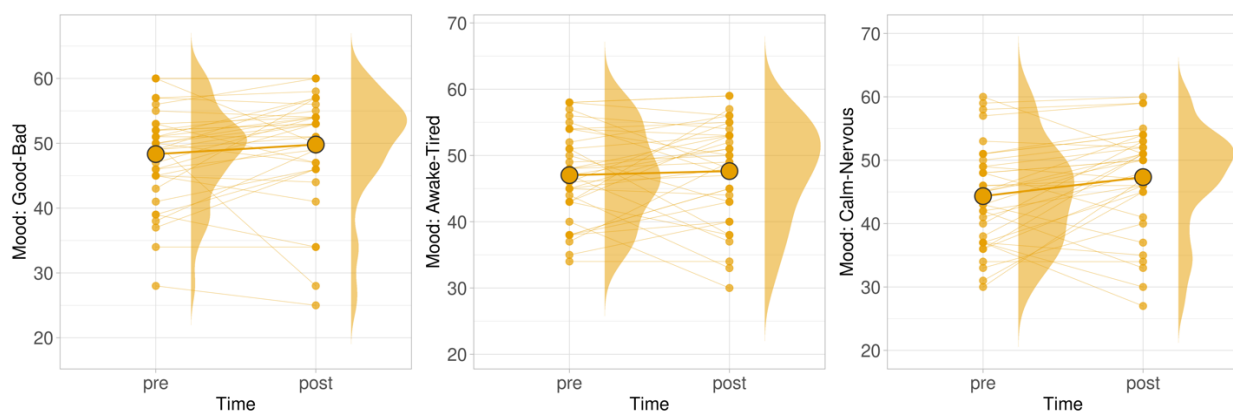

**Table 2.** Descriptive statistics of all outcome variables.

| Outcome Variable | | Mean (SD) | Theoretical Range |
|---|---|---|---|
| Usability | | 67 (21) | [0, 100] |
| Motion Sickness | | 22 (27) | [0, 235.62] |
| Engagement | | 4.2 (0.91) | [1, 5] |
| Spatial Presence | | | |
| Self-Location | | 3.8 (1.3) | [1, 5] |
| Possible Actions | | 3.7 (1.2) | [1, 5] |
| Social Presence | | 61 (22) | [0, 100] |
| Workload | | 2.8 (1.2) | [1, 7] |
| Reconnect | | 3.7 (0.79) | [1, 5] |
| Attitude | Pre | 3.2 (0.39) | [1, 5] |
| | Post | 3.1 (0.31) | [1, 5] |
| | Change | -0.052 (0.37) | [-4, 4] |
| Mood | | | |
| Good-Bad | Pre | 48 (7.4) | [10, 60] |
| | Post | 50 (8.5) | [10, 60] |
| | Change | 1.5 (6.3) | [-50, 50] |
| Awake-Tired | Pre | 47 (6.9) | [10, 60] |
| | Post | 48 (7.6) | [10, 60] |
| | Change | 0.61 (6.6) | [-50, 50] |
| Calm-Nervous | Pre | 44 (7.9) | [10, 60] |
| | Post | 47 (8.1) | [10, 60] |
| | Change | 3.0 (6.9) | [-50, 50] |

In analysis of the effect of cognitive impairment, we found significant differences in the two Spatial Presence subscales: Possible Actions ($\Delta$=1.07; SE=0.41; t(32.8)=2.60; p=0.014; 95% CI: [0.23, 1.90]) and Self Location ($\Delta$=1.17; SE=0.43; t(30.7)=2.72; p=0.011; 95% CI: [0.29, 2.05]). The participants with likely CI tended to report a higher Likeliness to Reconnect with partners, as well as a greater sense of Spatial Presence ("being there") in the VR environment (Figures 5 and 6).

**Figure 5.** Likelihood to Reconnect by Cognitive Impairment (at Left), Engagement by Cognitive Impairment (Middle), and Social Presence by Cognitive Impairment (at Right); Statistically Significant Differences Were Found Only for Likelihood to Reconnect

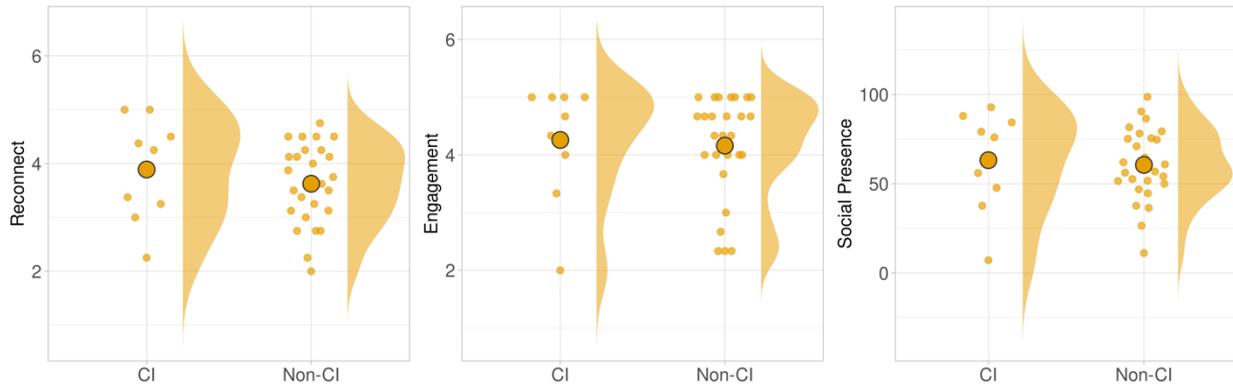

**Figure 6.** Possible Actions by Cognitive Impairment (at Left) and Self Location by Cognitive Impairment (at Right)

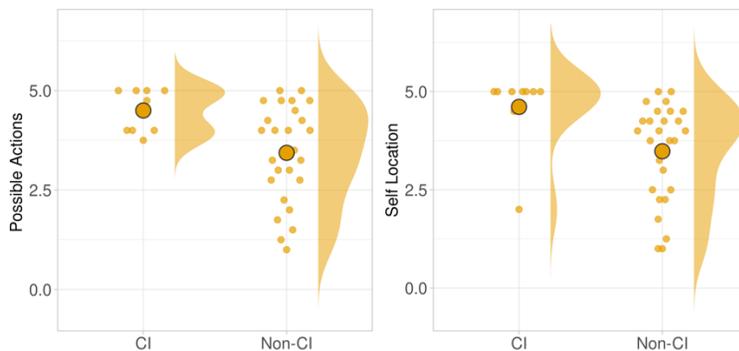

### 3.3. Similarity in Reconnect, Social Presence, and Engagement within Pairs (*RQ 4*)

We found small intra-cluster correlations for both Likelihood to Reconnect and Social Presence within the participant pairs (correlation coefficient $\rho$ = 0.08 and 0.11 respectively). This indicates that the effect of pairing had a small but statistically significant impact on these two variables. There was a somewhat larger intra-cluster correlation for Engagement with the VR environment ($\rho$=0.36). These findings make intuitive sense as positive social relationships are generally bi-directional, and engagement behaviors of a partner are likely to affect one's own sense of engagement.

### 3.4. The Impact of Participant Affect on Mood Changes, Social Outcomes and Engagement (*RQ5*)

Affective reactions to the virtual experience as measured by the SAM instrument were predictive of mood changes. The happy–unhappy dimension of the SAM was a predictor of changes in good–bad mood (b=-1.45; SE=0.71; t(32.0)=-2.05; p=0.049), as well as changes in awake–tired mood (b=-1.92; SE=0.72; t(32.0)=-2.67; p=0.012) and changes in calm–nervous mood (b=-2.01; SE=0.80; t(32.0)=-2.52; p=0.018). The excited–bored dimension of the SAM was a marginal predictor of changes in good–bad mood (b=-0.91; SE=0.50; t(32.0)=-1.85; p=0.075) and changes in awake–tired mood (b=-0.86; SE= 0.50; t(32.0)=-1.71; p=0.098). The dominance dimension of the SAM (in control / not in control) was not a significant predictor of changes in any of the mood dimensions.

The happy–unhappy dimension of the SAM was a predictor of Likeliness to Reconnect (b=-0.21; SE=0.09; t(32.0)=-2.33; p=0.027) as well as Engagement with the VR Environment (b=-0.22; SE=0.10; t(27.25)=-2.18; p=0.039). The excited–bored dimension of the SAM was a marginal predictor of Engagement (b=-0.13; SE=0.07; t(25.09)=-1.92; p=0.068). The dominance dimension of the SAM (in control / not in control) was not a significant predictor of Likeliness to Reconnect or Engagement.

## 3.5. Other Factors Predicting Social Outcomes and Engagement (*RQ6*)

We evaluated correlations between the primary study variables to construct a structural equation model for predictors of Engagement and Likeliness to Reconnect. Table 3 presents the correlation coefficients between variables in the model. Notably, we combined the two subscales of Spatial Presence in this analysis (by adding them together) due to a very high level of correlation between those subscales ($r=0.86$, $p<0.001$). The model that we derived is shown in Figure 7. Most of the paths, except the one from social presence to engagement were found to be significant, or marginally significant. The model has a chi-square of 1.884 (6, N=36); p=0.930; GFI=0.982; AGFI=0.937; CFI=1.000; RMSEA<0.001; 90% CI [0.000,0.063]; p[<=0.05]=0.943; SRMR=0.039), indicating a relatively good fit.

**Table 3.** Correlations between Variables Used for the Structural Equation Model

| | Reconnect | Engagement | Social Presence | Mood | Spatial Presence | Usability |
|---|---|---|---|---|---|---|
| **Engagement** | 0.79*** | | | | | |
| **Social Presence** | 0.19 | 0.35 | | | | |
| **Mood (Good–Bad) Change** | 0.41 | 0.53* | 0.15 | | | |
| **Spatial Presence (all)** | 0.43 | 0.57** | 0.41 | 0.46 | | |
| **Usability** | 0.23 | 0.45 | 0.23 | 0.24 | 0.36 | |
| **SAM (Happy–Unhappy)** | -0.36 | -0.38 | -0.13 | -0.35 | -0.28 | -0.55** |

Note: Mode: Spatial presence here is an additive combination of the "Possible Actions" and "Self Location" subscales and Mood (Good–Bad) is the Mood change before and after the experiment. After Holm correction, * p < 0.05; ** p < 0.01; *** p <0.005.

**Figure 7.** Structural Equation Model for the Main Study Variables; † p < 0.1; * p < 0.05; ** p < 0.01; *** p < 0.005

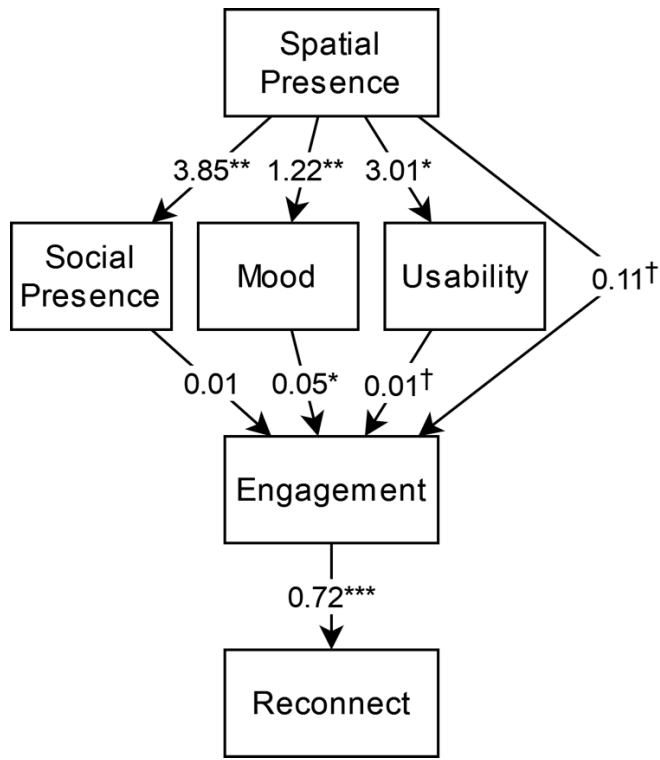

## 3.6. Qualitative Interview Analysis

We used a qualitative content-analysis approach (Elo & Kyngäs, 2008) to parse the participants' responses from the exit interviews. All responses were transcribed by a research assistant, who also developed a preliminarily coding schema based on commonly voiced themes related to the research variables. Two other research assistants independently reviewed the transcripts and coded the data segments by theme. After the research assistants completed their independent coding, they met to collaboratively resolve any disagreements about how the text segments were organized. The resulting themes and example quotations are presented in Table 4.

**Table 4.** Qualitative Content Analysis of Exit Interview Responses

| Theme | Sub-Theme Level 1 | Sub-Theme Level 2 | Number of Responses | Example Quotes |
|---|---|---|---|---|
| Affective Response | Positive | | 81 | "[VR] was a fun and exciting experience." (D013) |
| | Negative | | 10 | "A bit disappointing." (F008) |
| Mood Change | Bad Mood to Good Mood | | 8 | "Once I finally got the hang of it a little bit. Initially a bit awkward by clicking the wrong button and other things that were harder. Once I became more competent in it, I was able to do more, so it was less intimidating over time. When it was intimidating it was not so fun." (W013) |
| | Good Mood to Bad Mood | | 0 | |
| Likeliness to Reconnect | Likely | With Familiar People | 26 | "I would definitely use it for people that I do know." (D013) |
| | | With Strangers | 23 | "Yes I would use this tool for social interaction with other people, whether I know them or not." (W024) |
| | Unlikely | With Familiar People | 10 | "My daughter is in Philadelphia; that was who called me. So if she . . . I would rather be with her than in VR." (D002) |
| | | With Strangers | 9 | "For those I don't know . . . not sure if I would trust people I don't know to use this [VR] with." (W003) |
| Social Presence | Present | | 25 | "Well I would point out that if you were in a [real-world] location with another person where your conversation was no different . . . . It's just that the environment is created virtually in such a way as you feel that you are actually immersed in it and the communication is no different." (D008) |
| | Non-Present | | 24 | "If this is a study on how people develop friendships or socialization in virtual reality, then I don't understand how this would happen." (W028) |

| | | | |
|---|---|---|---|
| Spatial Presence | Present | 28 | "I like the idea that it's so three-dimensional that you really do feel like you are in a different place. Even though you realize it's imaginary it somehow feels realistic on a certain level so I like that part of it a lot." (D010) |
| | Non-Present | 12 | "I first felt limited by not seeing my own avatar, therefore I didn't see myself, and I do see myself in reality (my arms and legs). I also found walking to the next location rather strange." (W003) |
| Usability | Usable | 23 | "The ease of movement and the ease of moving objects." (D002) |
| | Unusable | 39 | "The headset needs to be more accommodating to an older person. Being able to work with glasses, and not being quite so heavy." (W013) |

One notable aspect of this data is that affective responses to the VR environment were predominantly positive (81 positive statements vs. 10 negative statements). Terms such as "comfortable," "fun," and "interesting" were common in the interviews. Such statements were most frequently linked to the experience of the 360-degree videos in the Travel module, and to a lesser extent with manipulating the photos in the Productive Engagement Module. A small number of participants expressed a negative reaction to the VR activities, mostly using terms that conveyed a sense of being overwhelmed, such as "confusing" and "nerve-racking." The presence of such negative experiences among the participants supports the view that VR may be experienced in very different ways by different individuals in the older adult population, and underscores the need for accessible and well-paced training activities. It is also notable that responses indicating mood changes were all in the direction of worse mood in the beginning of the VR sessions to better mood at the end of the sessions. These positive changes in mood were commonly linked in the interviews to feelings of increasing success or mastery in the VR environment.

Social presence and non-presence were frequent themes in the interview responses. For example, one participant (D008), a 73-year-old, male Ithaca resident characterized the VR environment as just another place to carry out desired socialization activities: "the communication is no different [in the VR]." This participant's conversation partner (F014), a 74-year-old, female Tallahassee resident shared a similar sentiment: "I would love to have access to use this [VR] to interact with others." In this case, both participants appeared to regard the VR as a "transparent" medium that did not interfere with reaching their socialization goals. Such participant pairing effects were notable in the exit interview data, as the researchers frequently found similar responses (positive or negative) expressed by both participants in a particular VR session.

Comments regarding the degree of social presence were mentioned in the interviews. Several of the participants, mentioned the inability to clearly read body language and gauge a partner's emotions as important issues in VR. In these statements about social presence, the inability of avatars to convey the full humanity of the other person was a predominant theme. Some participants had difficulty viewing the social VR activities as a real or complete human interaction. However, overall, the interview data indicated a fairly strong interest in using the VR platform again, especially if there was a lack of opportunities to visit someone in the physical world. Interest in using the platform to connect with new individuals (strangers) and with familiar individuals (friends and family) was expressed with approximately the same frequency.

Participants also mentioned the feeling of spatial presence as an important aspect of their VR social interactions. One participant (D001), a 67-year-old, female Ithaca resident with a likely CI , indicated that feeling immersed in the 360-degree videos contributed to her sense of sharing a social experience: "I felt I was eavesdropping on the groups seen and people moving around and I find that very exciting, trying to imagine what their conversations were and whether she got bit by a mosquito or did they see this." In this case, simply seeing other individuals represented in the environment (as part of the video) encouraged the participant to feel involved and prompted her to engage in "people watching" with her actual conversation partner. Another pair also expressed similar feelings of spatial presence and being strongly immersed in the environment. In contrast, several participants expressed frustration with not being able to move around voluntarily during the 360-degree video environments and said that they felt this reduced the sense of immersion. The inability to see one's own avatar was also frequently mentioned in the interviews as a technological obstacle to the feeling of spatial presence.

Although the usability ratings were high, the participants provided specific recommendations to enhance the system's usability and social effectiveness. One suggestion provided by the participants was to incorporate a survey-based partner-matching component in the system, to help ensure that individuals would be paired with amiable partners who shared common interests. Other recommendations were to allot more time for training and for completing tasks, and to use a wireless headset system. Issues with the headset were commonly mentioned in the interviews, including complaints about its weight, a lack of compatibility with eyeglasses, and feelings of awkwardness or discomfort with the unfamiliar equipment. There were also suggestions for the hand-held controller related to the size of the controls and the required finger dexterity, as well as the controller's complexity. These hardware accessibility issues are crucial considerations, and they indicate that the manufacturers of VR systems have not yet given much consideration to the needs of older adults in their product designs.

## 4. Discussion

One of the central findings of the current study is that the majority of our participants found the social VR environment to be a positive experience, with high ratings for Engagement and for Willingness/Likelihood to Reconnect with a social partner encountered in the VR environment. These findings are congruent with and lend further support to the limited number of studies that have previously evaluated older adult's responses to social VR environments (Baker et al., 2019; Roberts et al., 2019; Siriaraya & Ang, 2019). Similar to prior findings, we found that social VR can be rewarding for older adults, but also that aspects of the system, ranging from the pacing of activities, to the extent of environmental realism, to the physical design of hand controllers and other equipment, may need adjustment to better account for the needs of an older population. Our study also concurred with prior findings (Dermody et al., 2020; Huygelier et al., 2019; Kalantari et al., 2022a) in showing relatively low levels of simulator sickness and task workload among older adults engaging with VR.

While previous studies have found significant changes in mood states and in attitudes toward VR in various populations after exposure to the technology (Huygelier et al., 2019; Kalantari et al., 2022a), we did not observe such effects, except for in the single calm–nervous dimension of the *Multidimensional Mood State Questionnaire*. The reasons for this are uncertain, but it may be related to our participants' relatively high familiarity with information technology, as indicated in the Computer Proficiency instruments. While prior studies have mostly used participants who were entirely unfamiliar with VR technologies, our population's high level of technological savvy may have served to dampen the attitude and mood impacts of the exposure to VR during the experiment. Additionally, the prior studies that found mood effects were focused on restorative environments such as virtual gardens and were oriented explicitly toward the goal of mood improvement. Further, in this study exposure to VR was brief and limited to one experimental session.

Participants reported only a moderate high level of perceived Social Presence in the VR (an average rating of 61.21 out of 100). However, the variance on this measure was quite large (a standard deviation of 22.00). This was also reflected in our interviews, in which some participants were very enthusiastic about the ability to make social connections in the VR, while other participants were skeptical and indicated that it did not feel like a "real" human interaction. There are important technological features that can contribute to feelings of social presence, most notably in the ongoing development of more realistic avatars that can adequately mirror users' body language and expressions. Even so, the wide range of responses in our study would seem to indicate that some older adults may simply be more amiable to the context of virtual socialization compared to other older adult users. More research should be conducted to evaluate if these differences are related to specific user characteristics, personality features, or social goals.

Another notable finding in the study is that the measurements of Engagement, Social Presence, and Likeliness to Reconnect demonstrated a pair-wise relationship, in which participants who engaged with each other socially had similar responses. While these effects were not particularly strong, they show that positive and/or negative aspects of the social connection can build synergistically. We also found that affective responses to the environment—particularly the degree of reported pleasure and excitement—were correlated with Engagement and with Likeliness to Reconnect. Among other implications, this implies that designers should strive to integrate functionalities that produce pleasure and excitement in virtual social environments as a means of enhancing the social outcomes. Identifying appropriate content for this purpose is a complex task, as it entails creating balanced and refreshing experiences that remain "new" without being confusing or overwhelming. Fortunately, user-testing of VR environments and subsequent refinements of the content are relatively easy to implement (compared to real-world designs), and such testing will provide valuable feedback about affective response patterns.

Although the sample with a likely CI was small, we also evaluated the potential impact of having a CI on our response variables. For the measures of Engagement and perceived Social Presence, we did not find any impact of CI and the differences were found among those with and without a CI in Spatial Presence and Likelihood to Reconnect, were all positive and indicated that the environment is immersive for older adults with a CI and that the social engagements within the VR were rewarding for these individuals. These findings are congruent with the limited prior research on the acceptability of VR for older adults with mild cognitive impairment (Arlati et al., 2021; Kalantari et al., 2022a; Park et al., 2020).

Finally, the structural equation model that we developed to analyze mediation relationships among the primary study variables indicated that Spatial Presence is a key factor for predicting Engagement and positive social experiences (as measured by Likeliness to Reconnect). Mood change and Usability were identified as the most important variables mediating the relationship between Spatial Presence and Engagement, and Engagement was the variable that had the most impact on Likeliness to Reconnect (Figure 6). Our model indicates that efforts to make the VR environment more immersive (enhancing experiences of Spatial Presence) can have a tremendous impact on the response to a VR program and ultimately on the social outcomes. Numerous suggestions to help improve Spatial Presence emerged in our study, including better nuance, performance, and visibility of avatars (e.g., being able to see portions of one's own avatar body when looking around), and hardware improvements to reduce distraction (e.g., lightweight and wireless headsets).

One notable aspect of the structural equation model is that while Spatial Presence was a strong predictor of Social Presence which is congruent with the findings from Barreda-Ángeles & Hartmann (2022), but the link between Social Presence and Engagement was not significant. This is rather counterintuitive, as one would expect that the sense of encountering a "real" person in the VR (Social Presence) should be linked to Engagement and Likeliness to Reconnect. We suspect that this finding may be related to the very large variance of Social Presence scores among different participants, as discussed above. In any case, it is interesting that non-social aspects of the VR (Spatial Presence, Usability, aroused and positive Mood) appear to be driving Engagement and through that driving social outcomes, rather than the other way around. This finding has overlap with previous studies (Barreda-Ángeles & Hartmann, 2022; Jonas et al., 2019; McCreery et al., 2015; Siriaraya & Ang, 2012) that identified the sense of presence in VR as a central factor affecting the quality of social interaction.

### 4.1. Limitations and Future Research Directions

Although this was a pilot study sample size was relatively small which limited statistical power to and our ability to detect relationships. Of note, in this regard is our comparison of participants with likely CI vs. those without. The analysis conducted based on only 9 participants with likely CI. Future research would benefit from expanded sample sizes, particularly to allow for the evaluation of demographic variables. Our sample was primarily White, female and relatively well-educated with high technology skills. The current literature on social VR for older adults is extremely thin, and there is a great deal of room to analyze factors such as pairing effects, personality attributes, and various types of socioeconomic background variables. Longitudinal studies to evaluate the effects of social VR programs over time are also warranted.

The current study was also limited in by the VR program design. Technologically speaking, we used two commercial platforms and avatars developed by the researchers through those platforms. Some of the findings may not be fully generalizable to other VR engines. There are many types of emerging technological improvements in the realm of VR, particularly for avatars, that may be of interest to future research. Body-tracking and facial-tracking technologies have a tremendous potential to enhance Social Presence via more representative and fluid avatar dynamics (Baker et al., 2019).  Other aspects of our program design that could be improved are the variety of different activities available, and the amount of time allowed for participants to complete those activities. Designing and evaluating a variety of different types of social activities in the VR in future research will help to produce important data about types of programs that are valuable and of interest to older adult populations. A feeling of insufficient immersion time to become familiar with the system and to engage more fully with the social partner were common complaints in our exit interviews. The value of greater immersion time has to be balanced against the increasing risks of negative effects such as simulator sickness. Longitudinal studies across multiple immersion sessions may be an optimal way to evaluate the impact of increased exposure.

The extent of researcher moderation in VR social activities is also an important variable. Our study used a relatively heavy moderation approach, particularly in the first two modules, the research assistants were heavily involved in initiating the introductory conversations. Further, participants were assigned to a random social partner and did not have a selection choice. While a more open-ended social design might be preferable in some ways, it is important to think carefully about these parameters and to evaluate their impact, especially when working with potentially vulnerable populations such as individuals with CI. Obviously, there is a great deal of research needed in this area to determine the optimal means of balancing user autonomy against potential harms (not to mention moderator fatigue) in the service of achieving the social program's stated goals. Finally, the current study was limited to self-reported metrics to evaluate variables such as mood, affective response, and engagement. These survey instruments are valuable and effective research tools, but they may be usefully supplemented in future work with physiological data such as eye-tracking, heart rate, and EEG to provide additional information about how older adults experience social VR.

## 5. Conclusion

Prior research on older adults' experiences with social VR is extremely limited. The current study was conducted to evaluate the responses of older adults to a heavily moderated social VR program and to analyze features of the experience that may enhance engagement and positive social outcomes. We examined the participants' responses to the VR across multiple dimensions, including perceptions of spatial presence and social presence in the environment, changes in mood from before to after the VR session, changes in attitudes toward VR technology, the perceived usability of the interface, experiences of simulator sickness, task workloads, engagement levels in the environment, and social outcomes in terms of willingness/likeliness to reconnect with a social partner encountered in the VR. This in-depth, multi-dimensional analysis provides a valuable contribution to knowledge in the emerging study of older adults' experiences with VR. The results indicate that many older adults with and without CI find the technology to be enjoyable and usable, and that the extent of perceived Spatial Presence in the environment was a central driver of engagement and positive social outcomes in this population. The study also identified crucial areas for improvement that were important to our older adult participants, including the development of more realistic and responsive avatars, well-paced activities with sufficient training time, and controllers that are better suited for older adults. The need to understand factors relevant to older adults' experiences with social VR presents a tremendous potential for careful study and development.

**Data Availability Statement**

The datasets presented in this study can be found in online repositories. The anonymized quantitative and qualitative data, and R code is available via the following online OSF link: https://osf.io/4sxfn/.

**Acknowledgement**

This study was funded by the National Institute on Aging/National Institutes of Health (NIA 3 PO1 AG17211, Project CREATE IV – Center for Research and Education on Aging and Technology Enhancement).

## Appendix A. Descriptive Results for Engagement, Social Presence, and Reconnect for Each Survey Question

**Table A.1. Engagement (5-Point Likert Scale)**

| Question | Mean | SD | Min. | Max. | SE |
|---|---|---|---|---|---|
| I found today's experience to be engaging. | 4.69 | 0.67 | 2 | 5 | 0.11 |
| The experience I shared with my virtual partner today was meaningful. | 4.00 | 1.10 | 1 | 5 | 0.18 |
| The experience I shared with my partner today allowed me to express creativity. | 3.86 | 1.33 | 1 | 5 | 0.22 |

**Table A.2. Social Presence (Sliders on a 0–100 Scale, Reverse-coded)**

| Question | Mean | SD | Min. | Max. | SE |
|---|---|---|---|---|---|
| To what extent did you feel able to assess your partner's reactions to what you said? | 33.17 | 27.08 | 0 | 100 | 4.51 |
| To what extent was this like a face-to-face meeting? | 48.42 | 31.82 | 1 | 100 | 5.30 |
| To what extent was this like you were in the same room with your partner? | 33.89 | 28.20 | 0 | 100 | 4.70 |
| To what extent did your partner seem "real"? | 30.61 | 31.38 | 0 | 100 | 5.23 |
| How likely is it that you would choose to use this system of interaction for a meeting in which you wanted to persuade others of something? | 50.00 | 33.30 | 0 | 100 | 5.55 |
| To what extent did you feel you could get to know someone that you met only through this system? | 42.64 | 25.38 | 8 | 98 | 4.23 |

**Table A.3. Reconnect (5-Point Likert Scale)**

| Question | Mean | SD | Min. | Max. | SE |
|---|---|---|---|---|---|
| I generally liked the other participant. | 4.31 | 0.82 | 2 | 5 | 0.14 |
| I would be interested in getting to know the other participant better. | 3.89 | 0.98 | 2 | 5 | 0.16 |
| If given the chance, I would like to interact with the other participant again. | 3.86 | 1.13 | 2 | 5 | 0.19 |
| I could see myself becoming friends with the other participant. | 3.58 | 1.23 | 1 | 5 | 0.20 |
| The other participant generally liked me. | 3.61 | 0.69 | 3 | 5 | 0.11 |
| The other participant would be interested in getting to know me better. | 3.39 | 0.73 | 2 | 5 | 0.12 |
| If given the chance, the other participant would like to interact with me again. | 3.56 | 0.77 | 2 | 5 | 0.13 |
| The other participant could see himself/herself becoming friends with me. | 3.33 | 0.86 | 1 | 5 | 0.14 |

## Appendix B. Descriptive Results for Affective Response to

## Each Module

**Table B.1.** Results of the SAM Scale reported for each social VR module. Mean (SD).

|  | Introduction Module | Travel Module | Productive Eng. Module | Average |
|---|---|---|---|---|
| **Happy-Unhappy** | 2.5 (1.7) | 2.6 (1.8) | 1.9 (1.6) | 2.3 (1.4) |
| **Excited-Calm** | 4.4 (2.5) | 4.9 (2.4) | 4.7 (2.8) | 4.7 (2.1) |
| **Controlled-Dominant** | 5.0 (2.0) | 4.9 (1.9) | 5.4 (2.1) | 5.1 (1.5) |